\documentclass[useAMS,usenatbib]{mn2e}
\usepackage{graphicx,graphics,amsmath,amssymb}

\title[A transient component in the pulse profile of
PSR~J0738$-$4042]{A transient component in the pulse profile of
  PSR~J0738$-$4042} 
\author[Karastergiou et al.]
{A.~Karastergiou$^1$, S.~J.~Roberts$^2$, S.~Johnston$^3$, H.~Lee$^2$,
  P.~Weltevrede$^4$ and M.~Kramer$^5$\\
  $^1$Astrophysics, University of Oxford, Denys Wilkinson Building,
  Keble Road,
  Oxford OX1 3RH, UK\\
  $^2$Information Engineering, University of Oxford, Parks Road, Oxford OX1 3PJ, UK\\
  $^3$Australia Telescope National Facility, CSIRO, P.O. Box 76, Epping, NSW 1710, Australia\\
  $^4$Jodrell Bank Centre for Astrophysics, The
  University of Manchester, Alan Turing Building, Manchester, M13 9PL,
  United Kingdom\\
  $^5$Max-Planck-Institut f\"ur Radioastronomie, Auf dem Huegel 69, 53121 Bonn, Germany\\
} \date{\today}
\begin{document}


\pagerange{\pageref{firstpage}--\pageref{lastpage}} \pubyear{2008}

\maketitle

\label{firstpage}

\begin{abstract}
  One of the tenets of the radio pulsar observational picture is that
  the integrated pulse profiles are constant with time. This
  assumption underpins much of the fantastic science made possible via
  pulsar timing. Over the past few years, however, this assumption has
  come under question with a number of pulsars showing pulse shape
  changes on a range of timescales. Here, we show the dramatic
  appearance of a bright component in the pulse profile of
  PSR~J0738$-$4042 (B0736$-$40).  The component arises on the leading
  edge of the profile. It was not present in 2004 but strongly present
  in 2006 and all observations thereafter. A subsequent search through
  the literature shows the additional component varies in flux density
  over timescales
  of decades. We show that the polarization properties of the
  transient component are consistent with the picture of competing
  orthogonal polarization modes.  Faced with the general problem of
  identifying and characterising average profile changes, we outline
  and apply a statistical technique based on a Hidden Markov
  Model. The value of this technique is established through
  simulations, and is shown to work successfully in the case of low
  signal-to-noise profiles.

\end{abstract}


\section{Introduction}

The radio emission from pulsars is characterised by a range of dynamic
phenomena that take place on various timescales. Microstructure is
observed at the shortest ($\mu$s) timescales, stochastic or organised
changes in the pulse shape occur on timescales of the rotational
period $P$, while other phenomena such as nulling (e.g. Lorimer \&
Kramer 2005)\nocite{lk05}, where radio emission totally switches off,
may last significantly longer than one rotation.  An examination of
the average pulse properties of a given pulsar however, demonstrates
that the mean shape of a sufficiently large number of individual
pulses remains remarkably constant over long periods of time. It is
this quality of radio pulsars that makes them extremely useful tools:
knowing the exact shape of the pulse profile increases the precision
of the measurement of the pulse time of arrival, which is the basic
quantity in pulsar timing experiments.

\begin{figure*} 
\includegraphics[width=1.05\textwidth]{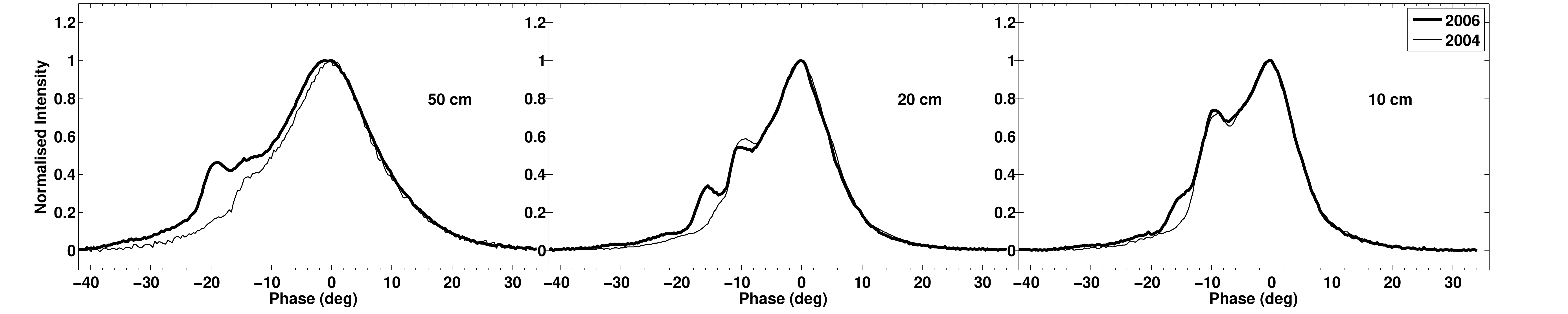}
\caption{The average profile of PSR J0738$-$4042 as observed with the
  50, 20 and 10~cm receivers at Parkes, in the first half of 2004
  (thin line) and the second half of 2006 (thick line). The
  change in the leading edge of the profile shape is visible at all
  frequencies.  \label{fig:bb}}
\end{figure*}
\begin{figure} 
\includegraphics[width=0.48\textwidth]{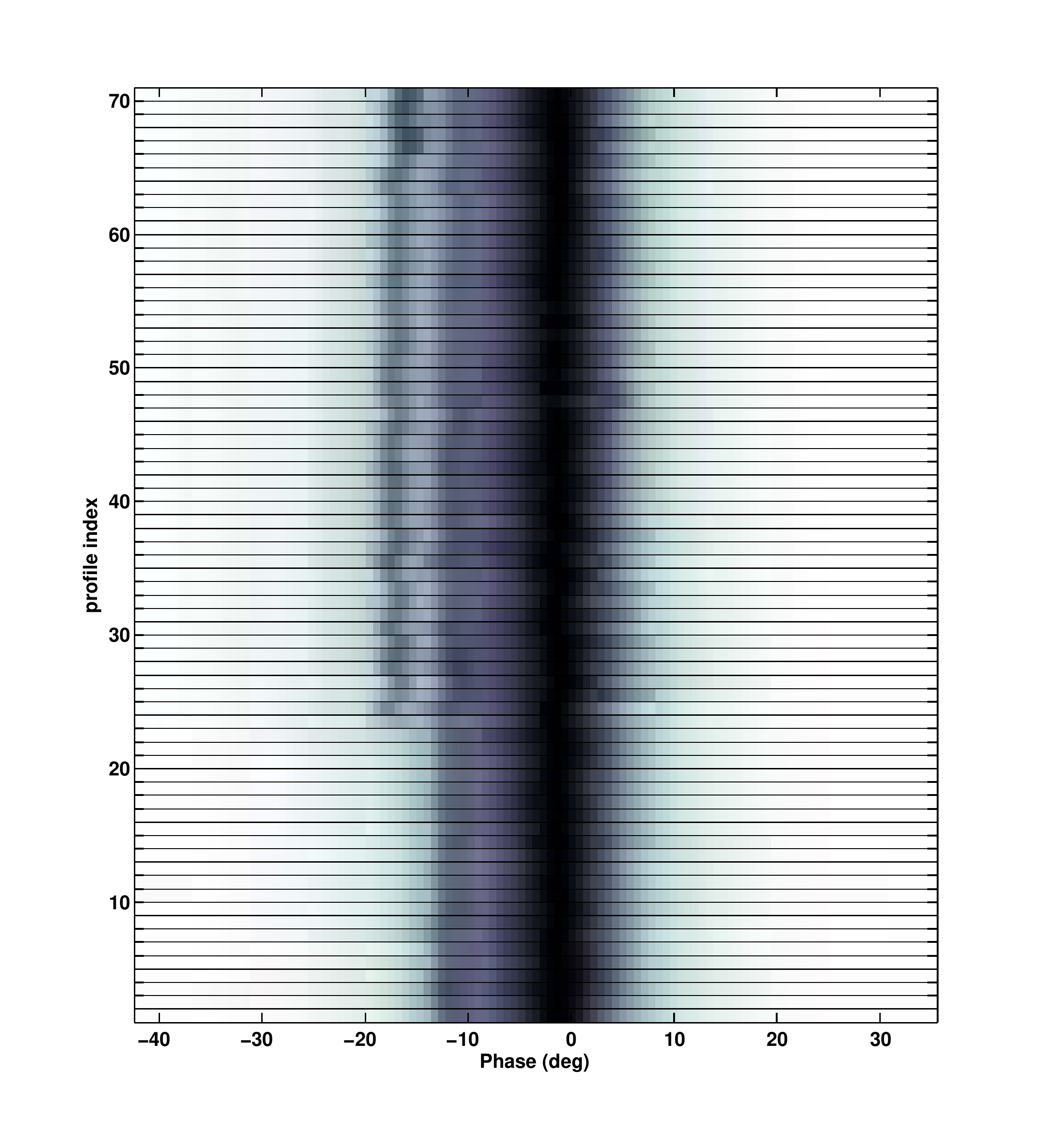}
\caption{A greyscale representation of the intensity of 71 profiles, observed at {\em irregular}
  intervals between 2003 and 2011. The peak intensity of each profile
  is normalised to unity; the profiles are aligned by cross-correlation with a top-hat function. The change
  occurs at profile 23.
  \label{fig:71profiles}}
\end{figure}

Although it is rare that the integrated profiles of pulsars change
with time, it is not unprecedented. Timescales of changes range from
hours through to months and years. On short timescales are the
rotating radio transients (RRATs) (McLaughlin et
al. 2006)\nocite{mll+06}, which are thought to be neutron stars that
only emit individual bursts of emission at irregular and infrequent
intervals. A group of pulsars show mode changing, where two distinct
and different pulse profiles are observed over timescales of hours
(e.g. Gil et al. 1994)\nocite{gjk+94}. The intermittent pulsar
PSR~B1931+24 (Kramer et al. 2006)\nocite{klo+06} is present for 5 to 10
days before its emission ceases for some 30 days. In this case, the
derivative of the rotational period, $\dot{P}$, changes between the on
and off phases and is therefore correlated with changes in the radio
pulse profile. Similar results associated with more subtle profile
changes are reported in Lyne et al. (2010), and another intermittent
pulsar (PSR~J1832+0029) is discussed in Kramer et
al. (2008)\nocite{kra08}. In PSR~J1119--6127 pulse changes were seen
which appeared to be associated with glitch activity (Weltevrede et
al. 2011)\nocite{wje10}.  There is strong evidence that all these
effects (nulling, mode changing and intermittency) are magnetospheric
in origin.

A further long-term effect, due to geodetic precession of a pulsar in
a binary orbit, can also cause pulse shape changes. These have been
observed in e.g. PSR~B1913+16 (Kramer 1998, Weisberg \& Taylor
2002\nocite{kra98,wt02}) and J1141--6545 (Manchester et
al. 2010)\nocite{mks+10}. Periodic changes in the average profile of
PSR~B1828$-$11 were interpreted as free precession by Stairs et
al. (2000)\nocite{sls00}, however free precession of solitary neutron
stars is considered unlikely (Sedrakian et al. 1999\nocite{swc99}) on
theoretical grounds.  Changes in the pulse profile due to precession
relate to changes in the viewing geometry rather than magnetospheric
effects.

The average pulse profiles of pulsars are partially linearly
polarized, to a lesser or higher degree. Highly energetic pulsars
feature high degrees of linear polarization (e.g. Weltevrede \&
Johnston 2008)\nocite{wj08b}. There is good evidence to suggest that
the observed emission results from the superposition of two
orthogonally polarized modes (OPM), which arise and propagate inside
the pulsar magnetosphere (e.g. McKinnon \& Stinebring 2000,
Karastergiou et al. 2001)\nocite{ms00,kkj+02}. Comparable intensities
of the modes has been favoured observationally as the cause for
reduced linear polarization in pulsars at typical ($\sim$1~GHz)
observing frequencies. A further observational consequence is that the
changes in the structure of the total power average profile with
observing frequency are often coupled with particular changes in the
degree of linear polarization, reflecting the spectral behaviour of
the orthogonal polarization modes (as discussed in Karastergiou et
al. 2005, Smits et al. 2006)\nocite{kjm05}\nocite{sse+06};as the OPMs
become more equal in strength, the total power increases and the
polarization decreases.

Pulsar timing models need to incorporate all known physical phenomena
(intrinsic to the pulsar or not) that affect the measured
times-of-arrival, in order to achieve a floor of sensitivity that
would enable the discovery of extremely weak components to the model,
such as gravitational waves (e.g. Hobbs et
al. 2009)\nocite{hob08}. Pulsar timing uses template matching and
relies on the average profile not varying with time. Any variability
in the profile adversely affect the timing model. In the following, we
present data from a significant change in the average pulse profile of
PSR J0738$-$4042. We show how polarization data reveal details about
the change, and discuss possible interpretations. We present a robust
statistical technique to characterise the change, and explore its
potential physical origins.

\section{The total intensity profile of PSR J0738$-$4042}
\begin{table*}
\caption{\label{tab:history} 40 years of average profiles from PSR
  J0738$-$4042.}  
\centering 
\small
\begin{minipage}{200mm}
\begin{tabular}{@{}llll} 
  \hline 
  Date   & Frequency & Component at -15$\degr$& Reference  \\ 
  \hline 
  $<$1970& 1720 MHz  &  Strong and discrete & Komesaroff et al (1970)\nocite{kmc70}\\                
  $<$1975& 1400 MHz  & Strong and discrete  & Backer (1976)\nocite{bac76}\\
  $<$1977& 631 MHz   & Shoulder to main pulse  & McCulloch et al.(1978)\nocite{mhma78}\\
  $<$1977& 1612 MHz  & Strong and discrete  & Manchester et al. (1980)\nocite{mhm80}\\
  1979  &  950 MHz   & Strong and discrete  & van Ommen et al. (1997)\nocite{vdhm97}\\ 
  1990  &  950 MHz   & Weak shoulder to main pulse & van Ommen et al. (1997)\nocite{vdhm97}\\ 
  1991  &  800 MHz   & Absent & van Ommen et al. (1997)\nocite{vdhm97}\\
 1996  &  1375 MHz   & Absent & unpublished \\
 1997  &  1375 MHz   & Absent & unpublished \\
 2004  &  1375 MHz  & Absent & Karastergiou \& Johnston (2006)\nocite{kj06}\\
  2004  &  3100 MHz  & Absent & Karastergiou \& Johnston (2006)\nocite{kj06}\\
  2005  &  8400 MHz  & Absent & Johnston et al. (2006)\nocite{jkw06}\\
  2005  &  3100 MHz  & Absent & Johnston et al.  (2007)\nocite{jkk+07}\\
  2006  &  1369 MHz  & Strong and discrete & Noutsos et al. (2009)\nocite{nkk+09}\\
  \hline 
\end{tabular} 
\end{minipage}
\end{table*}
PSR~J0738$-$4042 was one of the first radio pulsars discovered (Large
et al. 1968)\nocite{lvw68}. It has a high dispersion measure of
160.8~cm$^{-3}$pc, and its spin period of $P=375$~ms and a period
derivative of $\dot{P}=1.61\times10^{-15}$ places it within the bulk
of normal pulsars on the $P$-$\dot{P}$ diagram.  Its relatively high
flux density (80~mJy at 1.4~GHz) has made it a consistent observing
target over the 40 years since its discovery.  The most recent
polarimetric profiles of the pulsar over a wide range of frequencies
have been published in Karastergiou \& Johnston (2006), Johnston et
al. (2006) and Johnston et al. (2007)\nocite{kj06,jkw06,jkk+07}.
Average profiles in three separate observing bands taken in 2004 are
shown in Fig. \ref{fig:bb} (thin line).  The profile at 1.4~GHz shows
one bright component with a shoulder component on its leading edge, on
what appears to be a broad pedestal of emission on the leading and
trailing edge.  Subsequently, the pulsar was observed as part of a
program to measure accurate rotation measures (Noutsos et
al. 2009). The 1.4~GHz profile, taken in 2006, is shown as the thick
line in Fig. \ref{fig:bb}.  An additional component, $\approx15\degr$
earlier than the main peak, can be seen on the leading edge of the
2006 profile, which is almost entirely absent in 2004. This component
is present at all three observing frequencies.
\begin{figure} 
\includegraphics[width=0.49\textwidth]{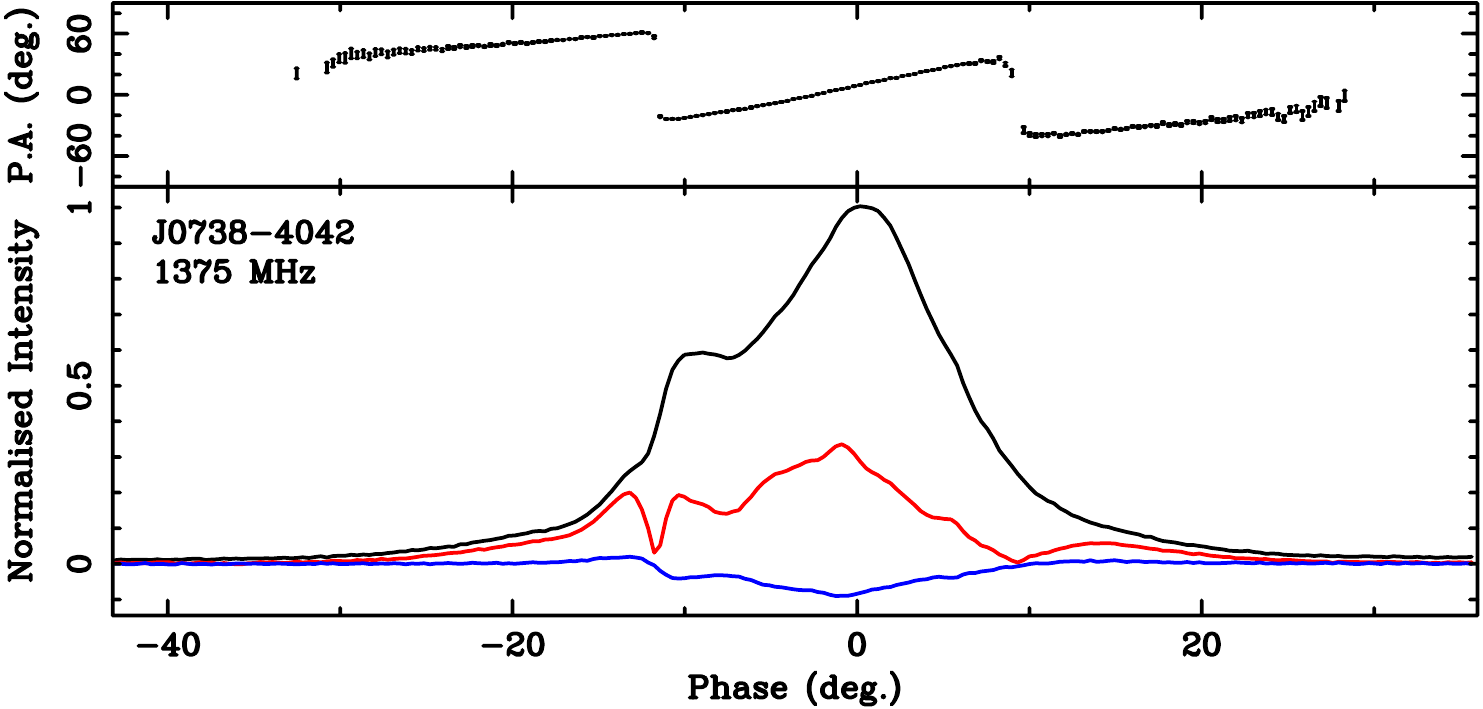}
\includegraphics[width=0.49\textwidth]{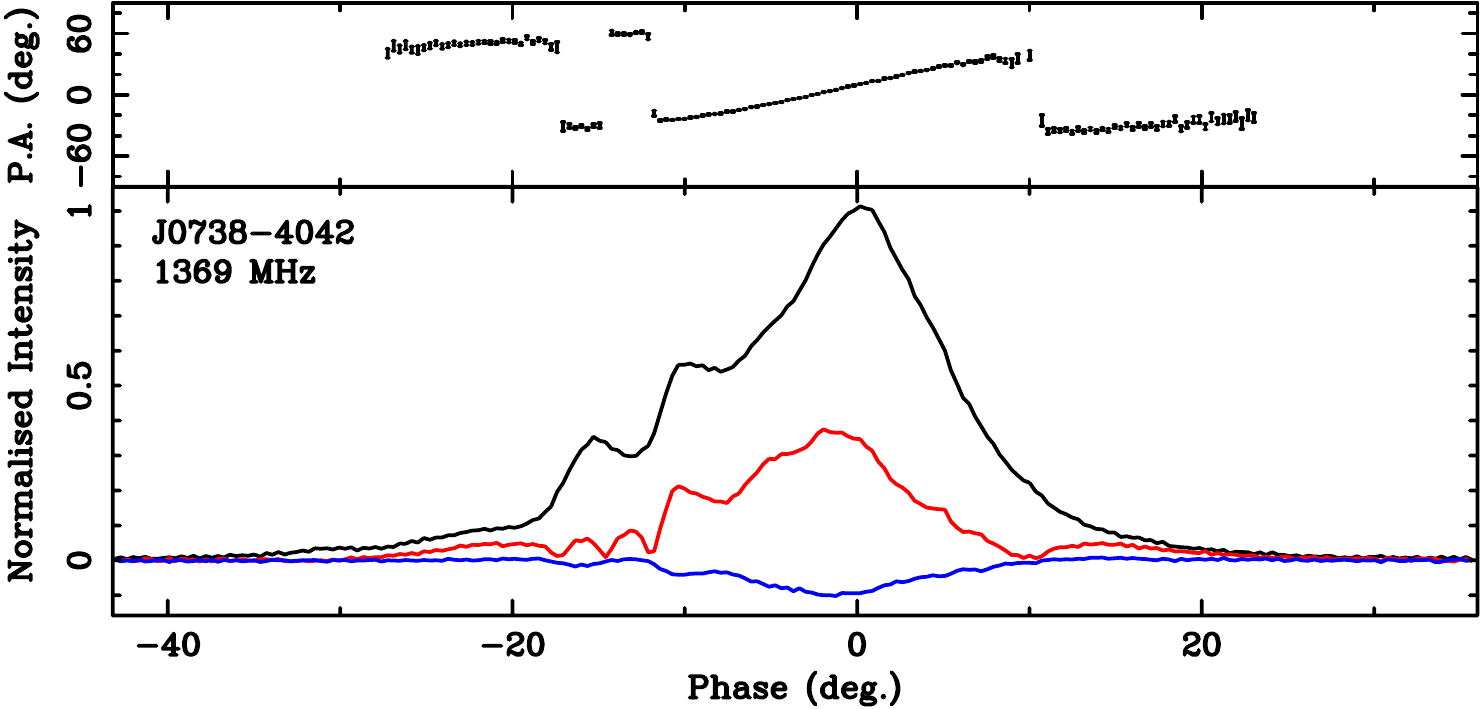}
\includegraphics[width=0.49\textwidth]{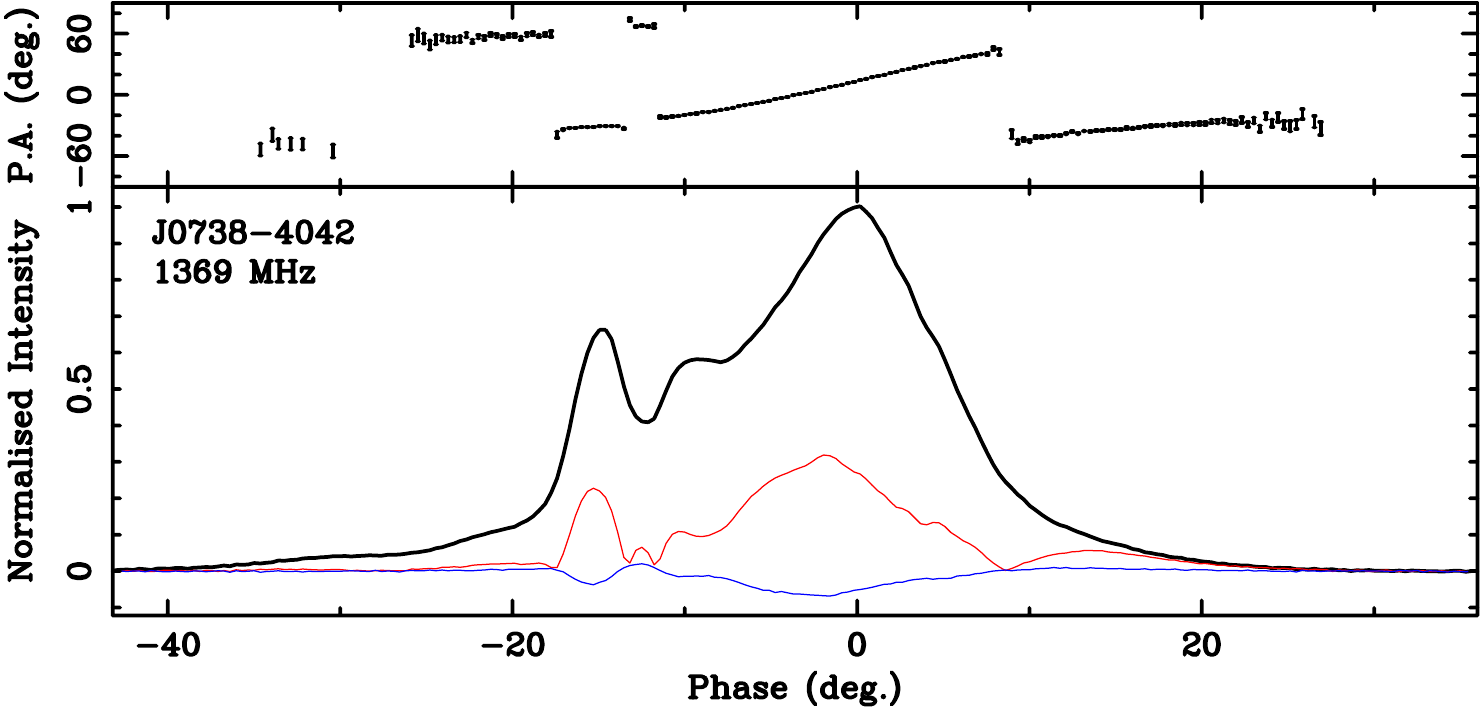}
\caption{Three average profiles of PSR~J0738$-$4042 in full
  polarization, from 2004, 2006 and 2010 (top to bottom). The total
  intensity(black), linear polarization (red) and circular
  polarization (blue) are shown just below the PA curve (dotted
  line). The extra component in the 2006 and 2010 profiles is clearly
  responsible for the observed additional orthogonal jumps in the
  leading edge of that profile. These jumps coincide in phase with
  local minima in the linear polarization.
  \label{fig:poln20cm}}
\end{figure}
\begin{figure} 
\includegraphics[width=0.48\textwidth]{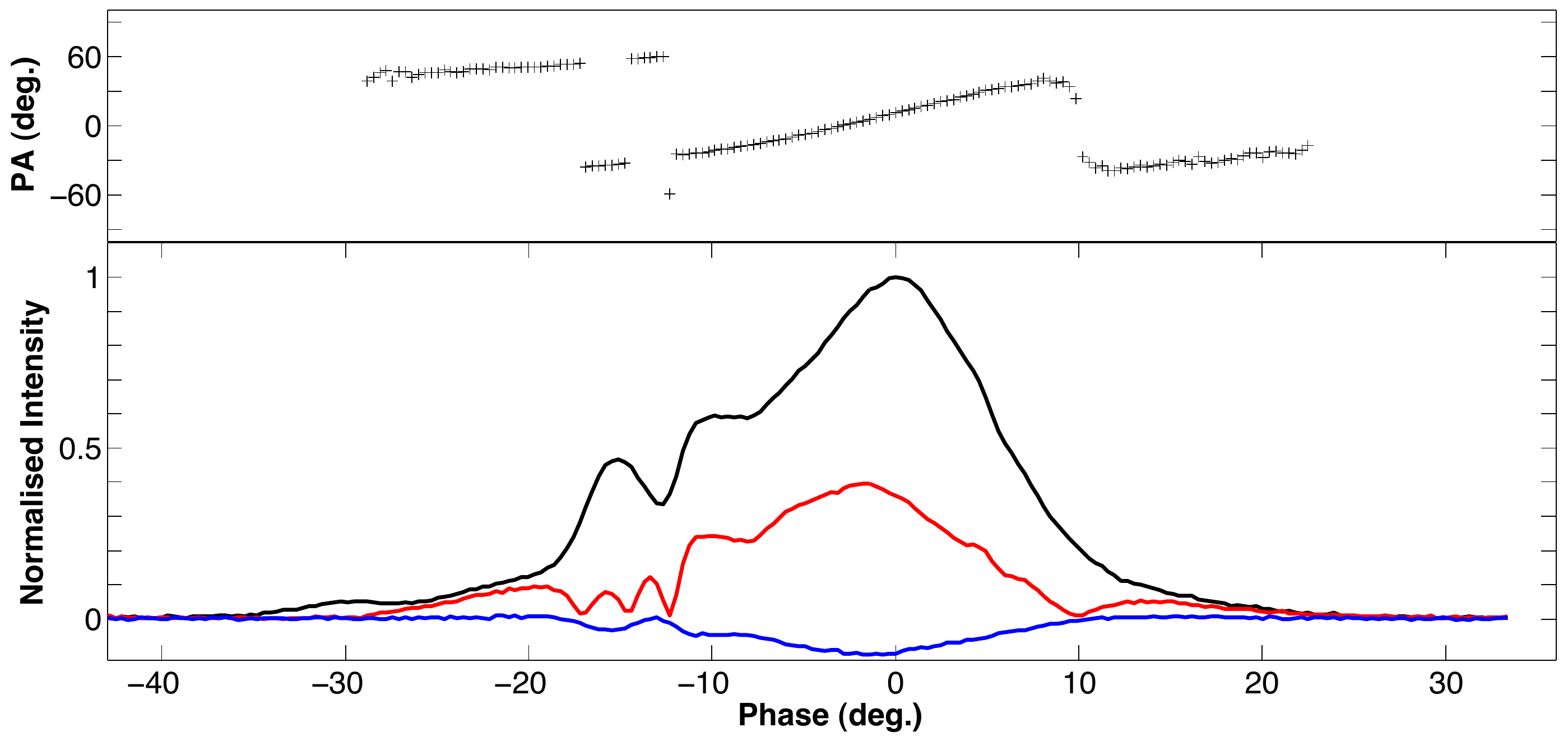}
\includegraphics[width=0.48\textwidth]{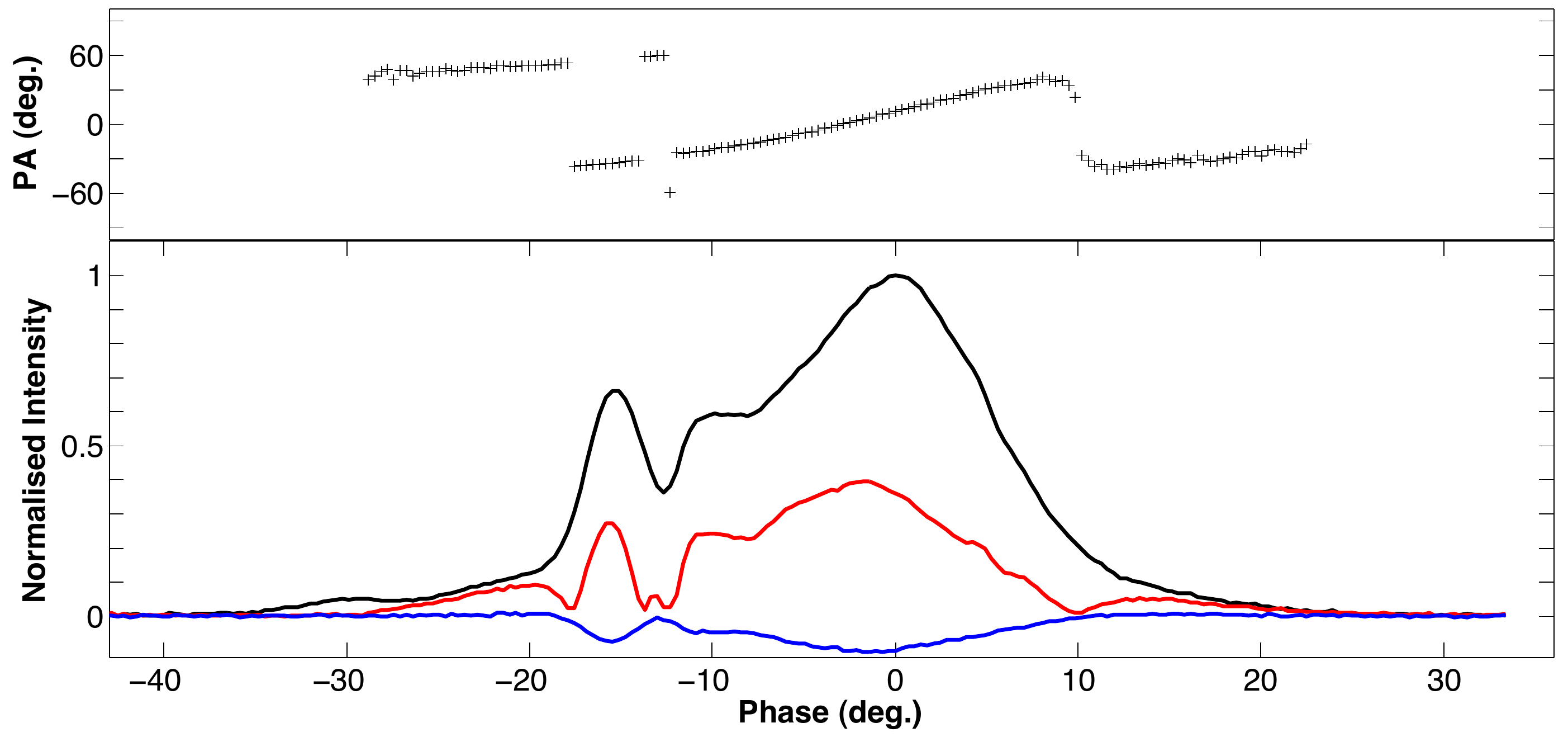}
\caption{Simulated polarization profiles resembling the data of 2006
  and 2010 from Fig. \ref{fig:poln20cm}. These profiles have been
  generated by adding an orthogonally polarized, Gaussian component of
  fixed width and variable amplitude to the polarization profile
  without the transient component (from 2004). Most features of
  Fig. \ref{fig:poln20cm} around pulse phase $-15\degr$ are well
  reproduced.
  \label{fig:polnmodel}}
\end{figure}

Armed with this result, we looked through the literature for other
published profiles of this pulsar.  Table \ref{tab:history} includes a
summary of available data, where the date and observing frequency are
given with a note relating to the presence of the leading
component. Two facts are immediately evident. First, there is a
period between 1991 and 2005 where the component is totally
absent. Secondly, the presence or absence of the leading component is
a broadband phenomenon; there are no contradictory observations at a
particular frequency. This is clearly seen in the representative, high
S/N profiles in Fig. \ref{fig:bb}, all showing additional leading edge
emission in 2006.

Figure \ref{fig:71profiles} shows data from 71 pulse profiles taken
with the 21~cm receiver of the Parkes telescope.  The profiles, which
are not collected at regular intervals, span a range of dates from
February 2003 to January 2011.  The index of each profile as shown on
the y-axis is used for the statistical analysis that follows. The
duration of each observation varies from 120 to 1199~s.  All profiles
in Fig. \ref{fig:71profiles} have been normalised to the peak
intensity and aligned by cross correlation with a top hat function.
Although absolute flux density calibration is not available for all
observations, an analysis of the S/N as a function of observing time
reveals that the peak flux density remains broadly similar, within a
$\approx20$\% margin.  The crucial change in the shape of the leading
edge can be clearly detected after profile \#23 with the ``new''
component remaining on all subsequent profiles.  Towards the end of
the sequence the new component has grown to its brightest amplitude.

\section{Changes in the polarization profile}\label{section:poln}
As mentioned in the introduction, polarization is a useful diagnostic
in the interpretation of pulsar radio emissions. Figure
\ref{fig:poln20cm} shows three average profiles of PSR J0738$-$4042,
from 2004, 2006 and 2010. It is immediately obvious that the
additional component which appears between 2004 and 2006 is also
associated with a different polarization state. Between pulse phase
$-20\degr$ and $-10\degr$, there are 3 orthogonal polarization jumps
in the 2006 and 2010 profiles, as opposed to a single jump in the 2004
data. It is also evident that the degree of linear polarization in
this region of the profile is related to the total power, and that
there are local minima directly related to the orthogonal PA jumps. A
comparison between the 2004 and 2006 data shows that as the total
power increases around pulse phase $-16\degr$, the linear polarization
drops.

We attempted to reproduce the observed changes in polarization using a
simple model. The starting point of the model is the polarization
profile obtained from observations in 2004 shown at the top of
Fig. \ref{fig:poln20cm}. To this, we added a Gaussian component
centred at phase $-15.47\degr$, which is 100\% polarized. We computed
the total intensity I$_c$ of the simulated component as:
\begin{equation}
{\rm I_c}(\phi) = A \exp \left [ -(\phi-15.47)^2/(2\sigma^2) \right ],
\end{equation}
where $A$ the amplitude and $\sigma$ the width of the Gaussian, and
$\phi$ the pulse phase. For each pulse phase bin, we set the
polarization of the simulated component to be orthogonal to the
polarization of the 2004 profile, by computing its Stokes parameters
(I$_c$,Q$_c$,U$_c$,V$_c$) relative to the Stokes parameters of the 2004
profile (I$_o$,Q$_o$,U$_o$,V$_o$), taking into account that:
\begin{equation}
\frac{\rm U_c}{\rm Q_c} = \frac{\rm U_o}{\rm Q_o},
{\rm I_c} = \sqrt{\rm Q_c^2 + U_c^2 + V_c^2}
\end{equation}
Therefore, Q$_c$, U$_c$ and V$_c$ can be computed for every pulse phase bin as:
\begin{equation}
{\rm Q_c} = -\frac{\rm Q_o I_c}{\sqrt{\rm Q_o^2 + U_o^2 + V_o^2}}
\end{equation}
\begin{equation}
{\rm U_c}= -\frac{\rm U_o I_c}{\sqrt{\rm Q_o^2 + U_o^2 + V_o^2}}
\end{equation}
\begin{equation}
{\rm V_c}= -\frac{\rm V_o I_c}{\sqrt{\rm Q_o^2 + U_o^2 + V_o^2}}
\end{equation}
 where the $-$ signs ensure that the polarization vectors are
antiparallel on the Poincar{\'e} sphere. Finally, we add the Stokes
parameters of the new component to the 2004 data to produce simulated
polarization profiles.

Figure \ref{fig:polnmodel} shows the best results of our model, which
can be compared directly to Fig. \ref{fig:poln20cm}. We find that we
can reproduce the data very closely by setting $\sigma=1.43^o$ and
varying $A$ between $\sim$0.25 for the 2006 data and $\sim$0.5 for the
2010 data. The fact that, at the relevant pulse phase region, the 2004
profile is almost entirely polarized, and the 2006 and 2010 profiles
can be simulated by adding a 100\% and orthogonally polarized
component leads to the conclusion that the polarization profiles with
the additional component are direct observations of the superposition
of two, 100\% polarized, orthogonal modes of emission.

\section{Statistical description of the profile evolution}
\subsection{Hidden Markov models}
Visual inspection of the profiles of Fig. \ref{fig:71profiles}
suggests that the profile evolution of PSR J0738$-$4042 can be
described by a single change of state, an assertion that is supported
by the polarization modelling described in the previous section. We
utilize a fully probabilistic realization of a hidden Markov model
(HMM) to automatically identify putative state changes in the data.
Hidden Markov models \citep{Rabiner89} have been widely used for
inferring latent changes in sequential data. Consider a sequence of
observations, $\mathbf{Y} = \{ \mathbf{y}_t \}_{t=1}^T$, $\mathbf{y}_t
\in \mathbb{R}^d, \forall t$. The distribution of an observation,
$\mathbf{y}_t$ is determined by a corresponding \emph{hidden} state,
$s_t \in \{1,\ldots,J\}$ and a state dependent observation
probability.

The hidden state at time $t=1$ is determined by a prior
state vector, $\boldsymbol{\pi} = [\pi_1,\ldots,\pi_J]^\top$, where
$\pi_j = p(s_1=j)$. The \textit{Markov property} implies that a hidden state $s_{t-1}$, $s_{t}$ depends
only on $s_{t-1}$ and not on those at time $t-2$ and before:
\begin{align}
  p(s_t|s_{t-1},\ldots,s_1) = p(s_t|s_{t-1}).
\end{align}
State transitions are jointly
determined by a transition matrix, $\mathbf{A} = [a_{ij}]$, where
$a_{ij} = p(s_t=j|s_{t-1}=i)$ and observation likelihoods.  An observation $y_t$
depends only on the corresponding hidden state, $s_t$ and we utilize here a state-dependent Gaussian density, such that the predictive distribution over observations, conditioned on the hidden state, is given by:
\begin{align}
  p(\mathbf{y}_t|s_t=j) = \mathcal{N} \left(\mathbf{y}_t ;
    \boldsymbol{\mu}_{j}, \boldsymbol{\Sigma}_{j} \right),
\end{align}
where $\boldsymbol{\mu}_{j}$ and $\boldsymbol{\Sigma}_{j}$ are a the mean
vector and a covariance matrix for the $j^{\rm th}$ state.

Inference in a HMM proceeds by evaluation of the posterior
distribution over the parameters, $\{ \boldsymbol{\pi}, \mathbf{A},
\boldsymbol{\mu}, \boldsymbol{\Sigma} \}$ as well as the posterior
distributions of the hidden state variables, $p(s_t|\mathbf{Y})$. A
two-stage maximum-likelihood algorithm is often adopted, such as the
Baum-Welch algorithm \citep{BaumETAL70}, a special case of the
expectation-maximisation (EM) algorithm \citep{DempsterETAL77}. The
most probable state-sequence can be found using the Viterbi algorithm
\citep{Rabiner89}. The maximum-likelihood approach, however, has major
limitations; most notably overfitting and inference of underlying
model complexity, such as determining the most probable number of
states. In order to address these limitations, we employ a fully
Bayesian approach, which exploits the tractable bounds of
\textit{variational Bayes} approximations \citep{JordanETAL99}. The
Bayesian methodology allows for uncertainty to be handled at all
levels of inference, making the approach ideal for analysis of smaller
data samples. Furthermore, the number of underlying states is inferred
automatically such that sequences without significant state changes
are modeled by a single state process. This deviates significantly
from maximum-likelihood methods in which the data is forced into a set
number of states.

As posterior state probabilities are inferred for each datum, we can investigate with ease not only the existence of state changes but also track the entropy associated with state determination at each point. The information entropy associated with the posterior over the states, conditioned on observation $\mathbf{y}_t$ is hence given as:
\begin{align}
 \mathcal{H}_t = -\sum_{j} p(s_t = j | \mathbf{y}_t) \log p(s_t = j | \mathbf{y}_t).
\end{align}
The advantage of utilizing entropy lies in the fact that increases in entropy will occur in locations in which a full state-change is not supported by the data. If the logarithm is to base two, then the entropy is returned in \textit{bits}, with 1 bit of entropy indicating the existence of a fully supported state change.
\begin{figure}
\includegraphics[width=0.48\textwidth]{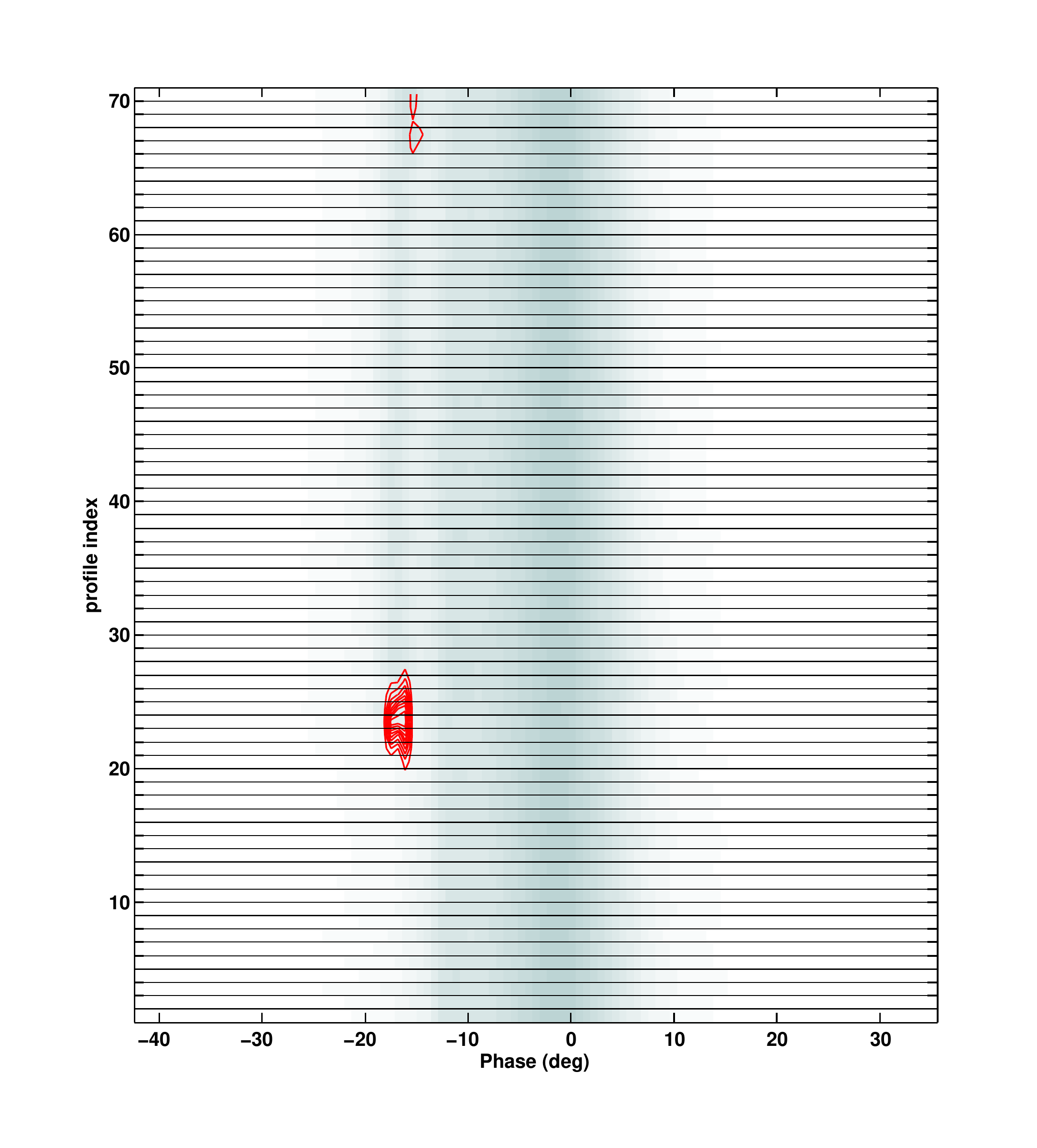}
\caption{Contours of the entropy of the HMM, superposed on the
  profiles from Fig. \ref{fig:71profiles}. The entropy is a measure of
  the probability of a state change, and reaches the value 1 in
  profile 23 and pulse phase -16. The contours are at entropy value steps of
  0.1, from 0 to 1.  Apart from profile 23, the probability of a state
  change is increased in profile 68.
  \label{fig:states}}
\end{figure}

We use the Bayesian HMM to infer the posterior distribution over the state sequence for each bin of the
pulse profile, using the total power amplitudes of each bin as our
observables. Figure \ref{fig:states} shows the entropy (in bits) of the state posterior
applied to all bins of all profiles, in contours superposed on the raw
data. The contours, which range from 0 to 1 in steps of 0.2, indicate
very high probability of a state change in profile 23, but also suggest
there may be further change in the most recent observations (profile
68). Artificially changing the order of the profiles moves the
identification point of the state change accordingly.

\subsection{Sensitivity analysis}
\begin{figure} 
\includegraphics[width=0.49\textwidth]{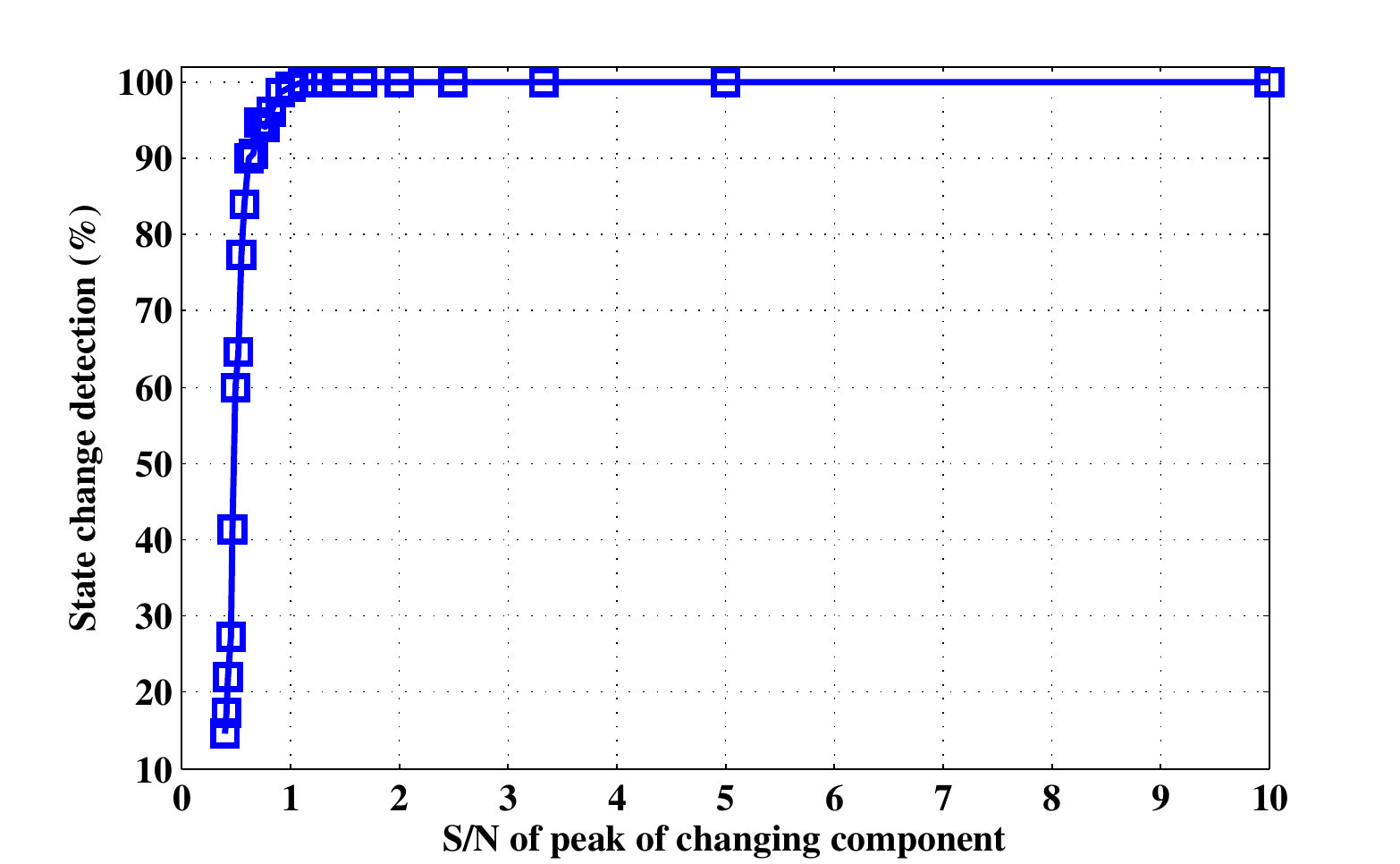}
\caption{Results of the sensitivity analysis. The curve shows the
  fraction of 1000 simulated profiles where the HMM successfully identifies a
  state change, versus the S/N of the new component. The HMM performs
  perfectly for S/N of 1 and above.
  \label{fig:s_analysis}}
\end{figure}
Although the HMM clearly and robustly identifies a state change in the
average pulse profile, it is not surprising that this analysis
performs well, given the large magnitude of the event (it can be
picked out clearly by eye). We have performed a sensitivity analysis
to test the performance of the HMM on noisier data. This involves
adding increasing levels of white noise to the profiles and comparing
the state-change identification with the original analysis. We have
performed this analysis by means of a Monte Carlo simulation,
producing 1000 versions of each profile with a given S/N and changing
the noise from 4 to 100 times the original in steps of 4
(i.e. $25\times10^3$ simulated profiles for each observed
profile). For each set of profiles at a given S/N, we run the HMM and
count the fraction of sets in which a state change is detected, as
well as the profile number that the transition occurs.

Figure \ref{fig:s_analysis} shows the results of the sensitivity
analysis, as the fraction of the 1000 sets where a state change is
detected, versus the S/N of the peak of the new component in the
simulated data.  Addition of the noise levels mentioned above
artificially decreases the average S/N of the peak of the new
component to values between 0.4 and 10. The HMM performs extremely
well from S/N per pulse phase bin of 1 and above.

\section{Discussion and conclusions}
We have shown that PSR J0738$-$4042 has undergone dramatic changes in
its average pulse profile in the period between 2004 and 2006.  The
changes affect the total amplitude and polarization of a single
component on the leading edge of the profile and are broadband at
least over the frequency range between 600 and 3100 MHz.  Examination
of the literature shows that the transient component was present
between 1970 and 1990, then absent until 2006 since when it has again
been present up to the current epoch (January 2011).  The
`intermittency' of this particular component can therefore be measured
in tens of years compared to e.g. the tens of days in PSR~B1931+24.
This serves as a unique example of magnetospheric changes on very long
time scales. This is further proof that there remains a lot to be
understood on the dynamic nature of pulsar magnetospheres on all
timescales.

The increase in intensity of the transient component is accompanied by
an orthogonal transition in the position angle of the linear
polarization, and by a decrease in the fractional linear polarization
of that part of the profile. When the extra component is absent the
leading part of the profile is almost completely polarized. Orthogonal
jumps in the polarization angle are common in pulsars with medium or
low levels of linear polarization, which implies that the degree of
polarization is affected by the presence of the two modes. In the
past, two models have been put forward to account for this behaviour:
the observed pulsar radiation either occurs in two partially polarized
orthogonal modes of emission which are emitted disjointly (e.g. Cordes
et al. 1978)\nocite{crb78}, or two entirely polarized orthogonal
modes, the superposition of which sets the total polarization (eg
McKinnon \& Stinebring 2000).  The data presented here strongly favour
the latter, as this is the first example where we see two distinct
states: a single mode and high polarization when the additional
component is absent, and low polarization associated with higher total
intensity when it appears. We show the validity of this interpretation
of the post-2006 data with a simple simulation.

Orthogonal polarization modes are thought to be related to propagation
effects within the pulsar magnetosphere (e.g. Melrose
2000)\nocite{mel00a}, which then strongly suggests a magnetospheric
rather than geometric origin for the observed profile changes. We
consider a geometrical effect such as free precession to be extremely
unlikely.  Magnetospheric effects have been shown to be responsible
for the profile changes seen in PSR B1931+24, and the correlations
between the period derivative and the profile shape changes of a small
number of pulsars in Lyne et al. (2010)\nocite{lhk+10} also point to
magnetospheric effects.  In PSR~J1119--6127, profile changes seem to
occur immediately following a glitch (Weltevrede et
al. 2010). Although there are insufficient existing data to trace the
behaviour of the timing of the pulsar over the long term, there is no
evidence for glitch activity between 2004 and 2006. Also, unlike
PSR~J1119--6127, a long series of single pulses from PSR~J0738$-$4042
taken after 2006 shows the transient component to be persistent in the
single pulses with no more than typical variability.

We have presented a robust statistical technique based on a Hidden
Markov Model that estimates the likelihood of state changes in the
total power data, and showed that this technique has great potential
for substantially noisier data in which similar systematic changes
occur.  We plan to apply the HMM technique to a large number of
pulsars for which regular observations have been conducted, to look
for and characterise statistically significant profile
variations. Most importantly, this technique provides a means of
correlating changes in the pulse profile with other physical
parameters, such as the period and period derivative, which could
prove extremely useful in partially accounting for non-Gaussian timing
residuals in pulsar timing models.
\section*{Acknowledgments}
AK is grateful to the Leverhulme Trust for financial support. The
Australia Telescope is funded by the Commonwealth of Australia for
operation as a National Facility managed by the CSIRO.

\bibliography{journals,modrefs,psrrefs,crossrefs,somerefs,hmm_bib}
\bibliographystyle{mn2e}
\label{lastpage}

\end{document}